\documentclass[preprint,numbers,sort,compress,12pt]{elsarticle}

\usepackage[utf8]{inputenc}
\usepackage{amsmath,amssymb,amsfonts,dsfont}
\usepackage{color}
\usepackage{mathtools}  

\usepackage{tikz}
\usetikzlibrary{shapes,arrows,positioning,calc}

\usepackage{subfigure}
\usepackage{tabularx,ragged2e,booktabs,caption}
\newcolumntype{C}[1]{>{\Centering}m{#1}}

\usepackage{bm}
\usepackage{textcomp}
\usetikzlibrary{arrows,shapes,chains}
\usepackage[english]{babel}
\usepackage[utf8]{inputenc}
\usepackage{algorithm}
\usepackage{algorithmicx}
\usepackage{algpseudocode}
\usepackage{amsmath}

\usepackage{multicol}  
\usepackage{multirow} 
\usepackage{fixme}
\fxsetup{status=draft}
\fxsetup{theme=color}
\usepackage{makecell}
\usepackage{amsmath} 
\usepackage{amssymb}  

\usepackage{diagbox}
\usepackage{textcomp}
\usepackage{longtable}
\usepackage{upgreek}
\usepackage[title]{appendix}

\usepackage{graphicx}

\algdef{SE}[DOWHILE]{Do}{doWhile}{\algorithmicdo}[1]{\algorithmicwhile\ #1}%

\usepackage{pdflscape}
\usepackage[breaklinks=true]{hyperref}
\usepackage{breakcites}

\usepackage{nomencl}
\makenomenclature
\usepackage{caption}
\usepackage{subcaption}
\usepackage{multicol}
\hypersetup{hidelinks}

\usepackage{amssymb}

\journal{Journal of Manufacturing Systems}

\begin{document}

\begin{frontmatter}



\title{An Efficient and Uncertainty-aware Reinforcement Learning Framework for Quality Assurance in Extrusion Additive Manufacturing}


\author[a]{Xiaohan Li\corref{cor1}}
\ead{xl617@cam.ac.uk}
\author[a]{Sebastian Pattinson\corref{cor1}}
\ead{swp29@cam.ac.uk}

\affiliation[a]{organization={Institute for Manufacturing, Department of Engineering, University of Cambridge},
addressline={Trumpington Street}, city={Cambridge},
postcode={CB2 1PZ}, 
country={UK}}

\cortext[cor1]{Corresponding author.}
\begin{abstract}
Defects in extrusion additive manufacturing remain common despite its prevalent use. While numerous AI-driven approaches have been proposed to improve quality assurance, the inherently dynamic nature of the printing process poses persistent challenges. Regardless of how comprehensive the training dataset is, out-of-distribution data remains inevitable. Consequently, deterministic models often struggle to maintain robustness and, in some cases, fail entirely when deployed in new or slightly altered printing environments. This work introduces an agent that dynamically adjusts flow rate and temperature setpoints in real-time, optimizing process control while addressing bottlenecks in training efficiency and uncertainty management. It integrates a vision-based uncertainty quantification module with a reinforcement learning controller, using probabilistic distributions to describe printing segments. While the underlying networks are deterministic, these evolving distributions introduce adaptability into the decision-making process. The vision system classifies material extrusion with a certain level of precision, generating corresponding distributions. A deep Q-learning controller interacts with a simulated environment calibrated to the vision system’s precision, allowing the agent to learn optimal actions while demonstrating appropriate hesitation when necessary. By executing asynchronous actions and applying progressively tightened elliptical reward shaping, the controller develops robust, adaptive control strategies that account for the coupling effects between process parameters. When deployed with zero-shot learning, the agent effectively bridges the sim-to-real gap, correcting mild and severe under- and over-extrusion reliably. Beyond extrusion additive manufacturing, this scalable framework enables practical AI-driven quality assurance across various additive manufacturing processes.
\end{abstract}


\end{frontmatter}



\section{Introduction}

Fused deposition modelling (FDM) is the most widely adopted extrusion-based additive manufacturing (AM) technique. It involves heating thermoplastic material and extruding it through a nozzle to construct objects layer by layer from a digital design. FDM offers significant design flexibility, facilitating the production of complex assemblies and supporting rapid prototyping \cite{praveena2022comprehensive}. As a result, it has found applications across various industries, including medical devices \cite{reinhard2024automatic}, aerospace \cite{froes2019additive}, architecture \cite{el2020critical}, and robotics \cite{zhang2022integrated}. Despite these advantages, FDM faces persistent challenges concerning quality consistency and production reliability. Defects such as geometric inaccuracies, mechanical weaknesses, and print failures can compromise the structural integrity and functionality of printed parts \cite{maurya2021investigation, mohan2021comparitive, baechle2022failures}. Even with precise calibration, extrusion inconsistencies often occur due to factors like filament slippage \cite{baechle2022failures}, nozzle blockages \cite{tlegenov2018nozzle}, filament quality variations \cite{zaldivar2018effect}, and motor-related issues such as overheating or mechanical wear \cite{prusinowski2017simulation}. Addressing these challenges is critical to improving FDM’s reliability and unlocking its potential for precision-driven applications.

Manual error correction in 3D printing remains labor-intensive, requiring considerable expertise. Human intervention is not only time-consuming but also inefficient, often resulting in suboptimal quality and increased material waste. To address these limitations, research has increasingly focused on continuous monitoring and real-time error compensation. Various sensors—including current \cite{guo2019fault}, inertial \cite{zhang2019deep}, acoustic \cite{li2022real}, and optical \cite{chen2024situ}—have been employed to monitor the printing process. Among these, camera-based monitoring stands out due to its flexibility, cost-effectiveness, and ability to detect common defects such as under-extrusion \cite{liwauddin2022continuous}, over-extrusion \cite{hsiang2020overview}, interlayer defects \cite{nguyen2024effect}, warping \cite{saluja2020closed}, and surface irregularities \cite{zhang2020detection}. These systems enable real-time error detection and correction, improving efficiency and reliability in the printing process. The integration of machine learning with closed-loop control systems has further advanced automated quality control in AM \cite{kim2024real,chen2024situ,naser2023automating,kim2022systematic}. However, effective decision-making remains a challenge for real-time error correction. Current approaches, such as those in \cite{brion2022generalisable, margadji2024iterative}, rely on basic parameter adjustments driven by computer vision predictions. For example, \cite{brion2022generalisable} employed multi-head neural networks to infer process parameters from large image datasets, while \cite{margadji2024iterative} refined predictions for mass production of a single object. Despite their effectiveness, these methods require extensive data and depend on simplistic decision-making processes, often leading to system instability and the introduction of new defects.

Reinforcement learning (RL) presents a promising alternative for dynamic parameter control. RL agents learn through interaction with their environment, optimizing complex relationships between process variables \cite{nian2020review}. Previous studies have demonstrated RL’s potential for AM process control. For instance, Ogoke et al. applied deep RL to adjust scan speed and laser power in laser powder bed fusion, but validation was limited to simulations \cite{ogoke2021thermal}. Similarly, Piovarci et al. developed an RL controller for direct ink writing but faced challenges in generalizing their model across multiple layers and varied environmental conditions \cite{piovarci2022closed}. Liao et al. explored RL for multi-material printing to optimize color control, but their system was restricted to managing filament feed rates \cite{liao2023learning}. These studies highlight the potential of integrating vision-based monitoring and RL but also expose key challenges. First, training an RL agent through direct interaction with a physical printer is prohibitively expensive due to material consumption and hardware wear. Existing RL controllers often depend on numerical models that struggle to accurately capture the complex dynamics of printing. Even simulations that approximate real-world behavior are typically limited to narrow conditions and lack scalability. Second, multiple sources of uncertainty in the printing process hinder the practical application of RL; without effective uncertainty management, the sim-to-real gap becomes very significant for real-world deployment. Third, current controllers often overlook the coupling effects and differing response curves between process parameters. Moreover, their design—whether in reward functions or training procedures—lacks a systematic, scalable, and structured approach, which is essential for practical applications.

In this paper, we propose an AI-driven framework for automating and optimizing setpoints in 3D printing, with a focus on addressing the aforementioned challenges. Our approach integrates the agent’s vision and RL modules through a vision-based probabilistic distribution, injecting quantified uncertainty into the decision-making process. This allows the agent to manage uncertainty effectively and make adaptive, informed decisions in real-time. Both modules are efficiently trained offline and deployed with zero-shot learning on a real 3D printer, resulting in a minor sim-to-real gap. The system achieves stable and accurate adjustments of flow rate and temperature, ensuring consistent quality. While validated in extrusion additive manufacturing, its scalability and adaptability make it a promising solution for broader applications across various AM processes.

\section{Methodology}

\subsection{Framework and novelty overview}

We propose a novel closed-loop control mechanism for 3D printing, incorporating quantified uncertainty directly into the decision-making framework. Specifically, the agent is divided into two interconnected modules: vision-based uncertainty quantification and RL control. The scaled distribution of vision analysis results serves as a quantified measure of uncertainty in a specific printing process segment. This distribution integrates the vision perception and RL control modules, enabling the agent to dynamically interpret process variations and make more informed, adaptive decisions in real-time. The generation of the scaled vision distribution is detailed in section \ref{ViT-based perception}, where we also outline the hard mask to extract key extrusion area. Section \ref{Deep Q-learning controller} presents the RL controller’s training process, elaborating its efficiency, uncertainty awareness, scalability, and structured methodology. Finally, Section \ref{Zero-shot 3D printer deployment} discusses the deployment strategy, including the key hyperparameters necessary for successful zero-shot learning in real-world 3D printing applications.

\begin{figure}[h]
\centerline{\includegraphics[width=0.6\columnwidth]{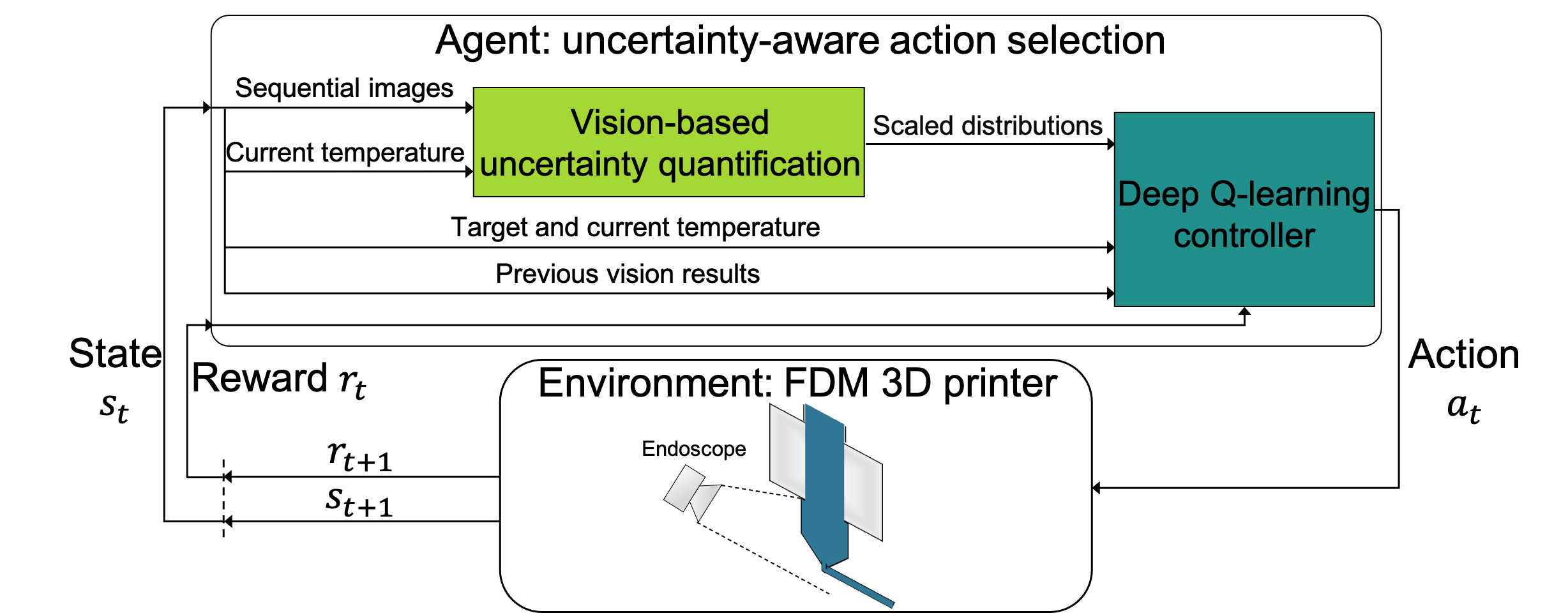}}
\caption{The general framework for 3D printing setpoint adjustment.}
\label{Fig1.framework}
\end{figure}

This framework, as illustrated in Fig. \ref{Fig1.framework}, is novel in several ways. First, it ensures efficient training for both modules, significantly reducing the need for costly real-world trials. Second, it equips the agent with uncertainty awareness, allowing it to make informed decisions under varying conditions. Third, it enhances interpretability by providing clear insights into how decisions are made. Finally, the reward function is designed with a structured and scalable approach that accounts for the coupling effects between multiple process parameters.

\subsection{Quantifying uncertainty in a printing segment}\label{ViT-based perception}

Three key factors are critical for effective uncertainty quantification in 3D printing vision monitoring. First, filtering out irrelevant information from the captured images is essential to ensure that vision analysis results are reliable and focused. This step minimizes distractions from non-essential features, allowing the system to concentrate on localized details—specifically, the exact extrusion amount near the nozzle tip. Second, the image analysis process must be highly robust during offline training. This requires the model to exhibit contextual awareness and be exposed to a diverse range of conditions. This capability enhances the model’s resilience to potential distribution shifts during actual 3D printing, ensuring consistent performance even under varying and unforeseen conditions. Third, it is vital to characterize uncertainty appropriately. This involves not only quantifying and representing uncertainty effectively within each printing segment but also structuring it in a way that can be efficiently simulated. This simulation-ready representation allows the RL controller to interact effectively, bridging the gap between simulation-based training and real-world deployment.

\subsubsection{Vision perception workflow and key area extraction}\label{Key area extraction}

\begin{figure*}[h]
\centerline{\includegraphics[width=\textwidth]{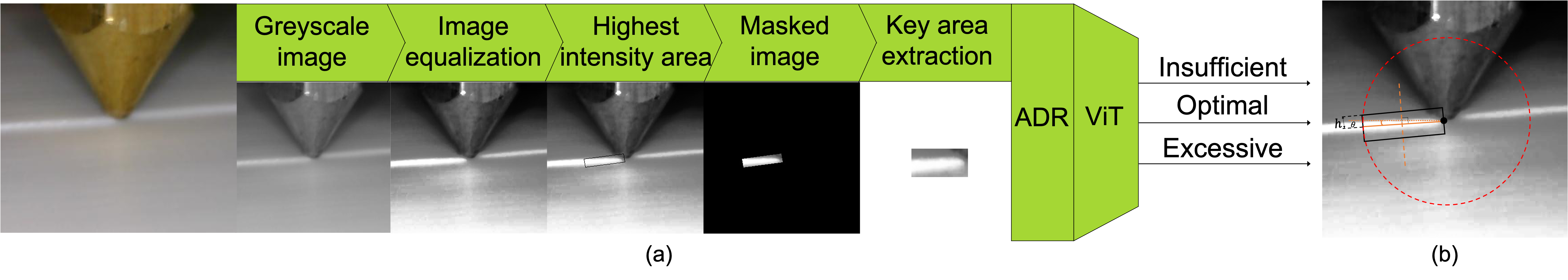}}
\caption{Vision perception: (a) workflow; (b) visual representation of key area detection.}
\label{Fig2.ViT_perception_method}
\end{figure*}

The first two factors are closely tied to vision analysis and are effectively addressed by the vision transformer (ViT)-based perception workflow, as illustrated in Fig. \ref{Fig2.ViT_perception_method}(a). The endoscope is mounted in a stable, fixed position relative to the nozzle, ensuring consistent imaging conditions throughout the printing process. Each captured image is cropped to $350\times 350$ pixels, focusing on the region immediately surrounding the nozzle, with the nozzle tip consistently centered in the image for accurate analysis. A hard mask is applied to isolate key extrusion-related regions, filtering out irrelevant information. These key areas are then augmented using automatic domain randomization (ADR), which involves random cropping (removing 10–30\% of pixels from the edges), horizontal flipping, and adjustments to brightness and contrast within a $\pm 10\%$ range. The random cropping step is particularly important as it simulates nozzle vibrations that naturally occur during printing, allowing the model to account for real-world variations. Each augmentation operation is applied independently with a 50\% probability, promoting diversity in the training data and enhancing model robustness. The augmented images, along with corresponding temperature data, are passed through ViT blocks to extract features. Leveraging ViT's contextual awareness to outperform traditional convolutional neural networks in robustness and adaptability. These features are then fed into a three-layer multilayer perceptron (MLP), which classifies each image into one of three extrusion categories: insufficient, optimal, or excessive. The classification task is relatively straightforward, as it primarily involves linking line thickness to three distinct extrusion classes. This simplicity is advantageous for practical applications of vision-based control, as it leads to more reliable and consistent results. In our framework, this reliability allows the vision system to focus on accurate detection, while the RL controller handles the more complex aspects of quality control. By sharing responsibilities in this way, the system ensures robust performance.

The necessity of using a hard mask to extract key extrusion areas, rather than relying solely on the ViT, lies in leveraging the unique characteristics of the printing and vision system setup. This targeted approach reduces the need for extensive data collection while minimizing classification errors in the vision analysis, ultimately improving efficiency and accuracy. The extraction of the key extrusion area follows a structured process designed to focus on the most relevant visual information. First, the nozzle tip image is converted to grayscale to simplify analysis and emphasize intensity variations. The freshly extruded material, located near the nozzle, is identified based on its higher pixel intensity—an effect caused by its proximity to the endoscope and increased light reflection. In contrast, regions on the opposite side of the nozzle display lower intensity due to shadowing effects. This contrast becomes even more pronounced after applying image equalization, which enhances brightness and sharpens distinctions between different areas. To accurately locate the key extrusion area, a line segment is defined with one end fixed at the nozzle tip (the center of the image). The opposite end rotates in 1-degree increments through a full 360-degree sweep around the nozzle. At each rotation angle, the average intensity along the line is calculated, and the line with the highest average intensity—corresponding to the freshly extruded material—is identified. The key extrusion area is then defined as a rectangular region, with the highest-intensity line serving as the horizontal midline. Two endpoints of this line are $s = (s_x, s_y)$ and $e = (e_x, e_y)$, and its perpendicular direction vector is given by
\begin{equation}
    \mathbf{g}=[s_y-e_y, e_x-s_x]^T.
\end{equation}
To determine the four vertices of the rectangle, denoted as $v_1$, $v_2$, $v_3$, and $v_4$, a distance $h$ is specified as half of the rectangular height. Specifically, 
\begin{equation}\label{eq: vertice 1}
    v_{1}=[s_x-\overline{\textbf{g}}_0h, s_y-\overline{\textbf{g}}_1h],\quad 
    v_{2}=[s_x+\overline{\textbf{g}}_0h, s_y+\overline{\textbf{g}}_1h],
\end{equation}
\begin{equation}\label{eq: vertice 2}
    v_{3}=[e_x-\overline{\textbf{g}}_0h, e_y-\overline{\textbf{g}}_1h],\quad 
    v_{4}=[e_x+\overline{\textbf{g}}_0h, e_y+\overline{\textbf{g}}_1h],
\end{equation}
where $\overline{\mathbf{g}}=\mathbf{g}/\|\mathbf{g}\|$ is the normalized direction vector of $\mathbf{g}$. Fig. \ref{Fig2.ViT_perception_method}(b) visually illustrates this computation, with the black dot at the image center representing the nozzle tip location. The red solid line indicates the maximum intensity line segment identified by rotating $\theta$ within the range of $[0^\circ, 360^\circ]$ along the dashed circle. The red dashed line represents the perpendicular direction vector $\mathbf{g}$. This systematic approach ensures that the analysis remains focused on the most relevant region of interest, reducing unnecessary computational overhead while improving the accuracy of extrusion classification. By combining this hard-masking technique with sufficient ADR and the ViT’s contextual awareness, the system achieves robust performance while minimizing data requirements.

\subsubsection{Scaled distribution of vision results}\label{Scaled vision perception distribution}

The third factor is addressed by leveraging the scaled distribution of vision results within an image window $w$, rather than relying on single-image outcomes, to provide uncertainty quantification in this printing segment. The extrusion condition is characterized by two key factors: the nozzle temperature measurement and the vision classification distribution. While the nozzle temperature is directly measured, the vision distribution is calculated as the proportion of classifications within the image window. Specifically, for $w$ results, we count occurrences of each extrusion category: $w_0$ for insufficient, $w_1$ for optimal, and $w_2$ for excessive extrusion. The distribution of these three classes is represented as $\mathbf{p} = [p_0, p_1, p_2]^T$ which is
\begin{equation}\label{eq:classification_probability}
    p_i =w_i/w,\quad \text{for}\quad i=0,1,2,
\end{equation}
where $\sum_{i=0}^2 w_i = w$ and $\sum_{i=0}^2 p_i=1$. The distribution of vision classification results serves as the foundation for the RL controller's decision-making process. For the controller to effectively learn flow rate and temperature adjustments, the environment must simulate these distributions sequentially, one printing segment at a time, while incorporating sufficient randomness.

The quality of these simulated distributions directly affects the sim-to-real gap. The more accurately these distributions are simulated, the smaller the gap will be when transferring the learned policy to real-world printing. To improve this simulation fidelity, we normalize the classification probability vector $\mathbf{p}$ by applying the transformation  $\mathbf{p} \gets \mathbf{p}/\max(\mathbf{p})$. This normalization scales the highest probability class to 1, with the remaining classes adjusted proportionally within a range of 0 to 1. This highlights the most likely classification for the current printing segment while preserving the relative uncertainty across classes. Normalizing the maximum probability class to 1 facilitates the mathematical representation to capture the key characteristics of both correct and incorrect classification distributions. A stochastic simulation environment is then constructed to replicate this normalized probability distribution $\mathbf{p}$. This environment is designed to present the RL controller with sufficient scenarios of both correct and incorrect classification distributions, further aligned with the precision level of the vision system, the details of which are provided in section \ref{Deep Q-learning controller}.

\begin{figure*}[!t]
\centerline{\includegraphics[width=\textwidth]{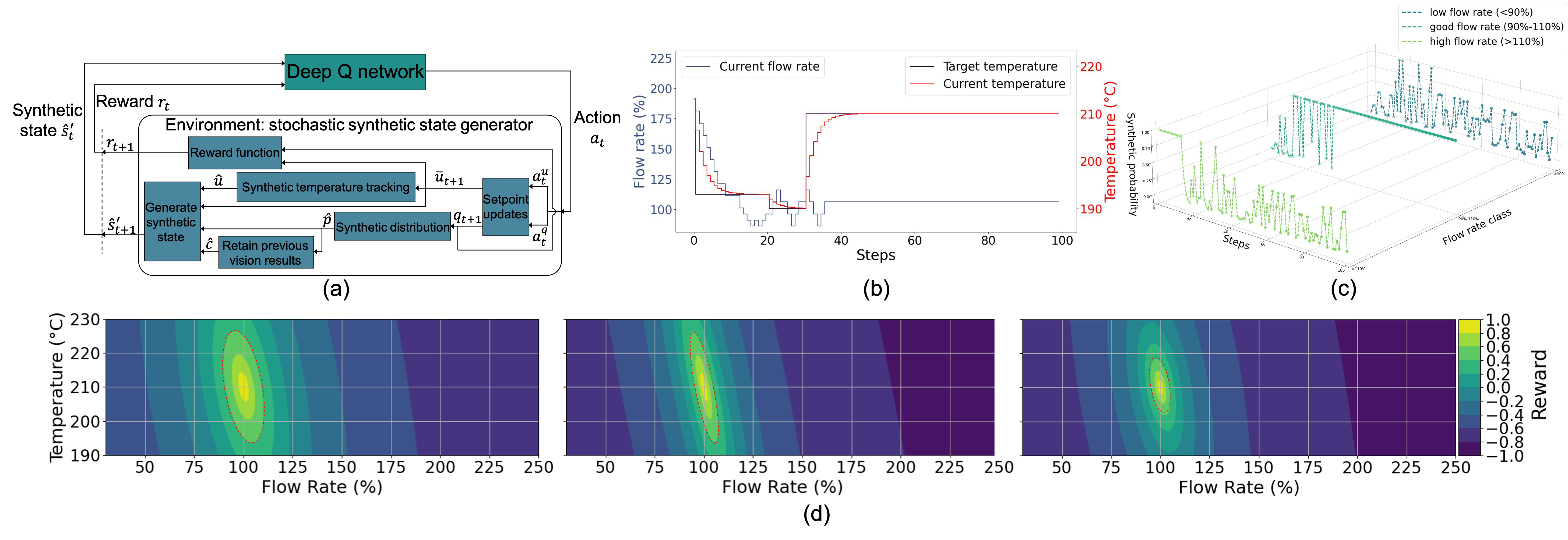}}
\caption{Stochastic simulator and synthetic data overview: (a) framework of the stochastic simulator; (b) synthetic temperature tracking and flow rate variations with asynchronous actions; (c) flow rate classification probabilities with exponential decay and Gaussian white noise; (d) reward shaping phases: phase 1, 2, and 3 from left to right, with phase 2 halving the minor axis and phase 3 halving both axes.}
\label{Fig3.deepQ_method}
\end{figure*}
\subsection{Deep Q-learning controller}\label{Deep Q-learning controller}

Although deep Q-learning is model-free, it demands extensive interactions with the environment. The exploration-exploitation trade-off also poses a significant challenge, as it may lead to suboptimal or even unsafe actions during the learning process. To mitigate these challenges, a stochastic simulation environment is designed to sequentially generate synthetic transitions, representing one printing segment after another. This setup effectively mimics the progression of real-world 3D printing, allowing the agent to experience diverse scenarios and uncertainties in a controlled setting. It allows flexible model architecture adjustments, facilitates validation of the reward function, and ensures that the model receives sufficient training before deployment on a real printer. Section \ref{Synthetic transitions and deep Q-learning training} details the generation of synthetic transitions and their role in replicating real-world uncertainties. Section \ref{Reward shaping for improved convergence} introduces a structured four-phase training process for the deep Q network (DQN), designed to progressively build the controller’s ability to achieve steady-state convergence and manage uncertainty.

\subsubsection{Synthetic transitions}\label{Synthetic transitions and deep Q-learning training}

 In 3D printing, step motors and heaters respectively control flow rate and nozzle temperature, each with different response times. Firmware settings are adjusted immediately, but the actual flow rate and temperature change gradually, with temperature response being slower. Therefore, flow rate and temperature actions are executed asynchronously, with one temperature action for every $\lambda$ flow rate actions. The optimal $\lambda$ is determined experimentally for timely corrections. At each time step $t$, five discrete actions are available for flow rate $a^q_t$ and temperature $a^u_t$. In specific, the action space $\mathcal{A}$ is 
\begin{equation}
    \mathcal{A} = \left\{ a_t = (a_t^q, a_t^u) \mid
    \begin{aligned}
         a_t^q &\in \{0\%, \pm 5\%, \pm 10\%\}, \\
        a_t^u &\in \{0^\circ \text{C}, \pm 10^\circ \text{C}, \pm 20^\circ \text{C}\}
    \end{aligned} 
    \right\}.
\end{equation}
When a temperature action $a_t^u$ is taken, the target temperature is updated as $\bar{u}_{t+1}=\bar{u}_t+a_t^u$. In an ideal scenario, temperature tracking follows the PID response curve of the nozzle heater. However, in the simulation, it is modeled using a synthetic response curve for computational efficiency. The current temperature is adjusted incrementally at each interaction, with the update proportional to the difference between the current and target temperatures. As the temperature approaches the target, the increments gradually decrease, ensuring a smooth convergence. This is formulated as
\begin{equation}
    \hat{u}_{t+1} \gets \hat{u}_t + (\bar{u}_{t+1}-\hat{u}_t)/\delta,\label{eq:temperature_track}
\end{equation}
where $\delta$ controls the rate at which the temperature converges to the target. As discussed in section \ref{Scaled vision perception distribution}, the scaled distribution $\mathbf{p}$ capture the dynamics of 3D printing. In synthetic transitions, the approximation $\hat{\mathbf{p}}=[\hat{p}_0, \hat{p}_1, \hat{p}_2]$ is generated using a noisy exponential decay function centered at the selected class $x_0 \in [0, 1, 2]$, corresponding to insufficient, optimal, and excessive extrusion respectively. This formulation ensures that the distribution retains its probabilistic structure while incorporating realistic variability, making it a suitable representation of uncertainty in extrusion conditions. Specifically,
\begin{equation}
    \hat{p}_i = 
\begin{cases} 
[\exp{(-\alpha \left| i-x_0 \right|)} + \mathcal{N}(0,\sigma^2)]^1_0 & \text{if } i \neq x_0, \\
1  & \text{if } i = x_0, 
\end{cases} \label{eq:hat_pi}
\end{equation}
where $\alpha$ controls the speed of exponential decay, $\mathcal{N}(0,\sigma^2)$ is a Gaussian white noise added to the classes that are not chosen. While $\hat{p}_i=1$ for the chosen class, other synthetic values are clipped into the range from 0 to 1 using the clamping function  \( [\cdot]_0^1 \). The notation for the classification result can be extended from the current classification result $x_0$ to previous $i$th classification results $x_i$ for $i=1,\cdots,\eta$. The previous $\eta$ classification results are retained in the vector $\hat{\mathbf{c}}=[\hat{c}_1,\cdots,\hat{c}_\eta]$, which are calculated as
\begin{equation}\label{eq:c}
    \hat{c}_i=x_i/n_c\quad\text{for}\quad i=1,\cdots, \eta,
\end{equation}
where $n_c=3$ is the class amount. At the initial time step, all $\eta$ values in $\hat{c}$ are initialized to represent ideal extrusion, set to $1/n_c$. As new classification results become available, $\hat{c}$ is progressively updated from $i=1$ to $\eta$, incorporating the evolving extrusion conditions. As illustrated in Fig. \ref{Fig3.deepQ_method}(a), the synthetic state $\hat{\mathbf{s}}^{'}_t \in \mathds{R}^{\eta+5}$ encapsulates prior classification outcomes $\hat{\mathbf{c}} \in \mathds{R}^{\eta}$, the current classification distribution $\hat{\mathbf{p}}\in\mathds{R}^3$, temperature $\hat{u} \in \mathds{R}$, and target temperature $\bar{u} \in \mathds{R}$. Specifically,
\begin{equation}\label{eq:synthetic_state}
    \hat{\mathbf{s}}^{'}_t= [\hat{c}_{\eta},\cdots, \hat{c}_1, \hat{p}_0, \hat{p}_1, \hat{p}_2, \hat{u}_t, \bar{u}_t ].
\end{equation}
Once simulated temperature tracking and scaled vision distributions are established, the transition at time step $t$ is defined by $\{\mathbf{s}^{'}_t, a_t, r_t, \mathbf{s}^{'}_{t+1}\}$ where the current action $a_t=(a^q_t,a^u_t)$ is from the admissible set $\mathcal{A}$ according to the synthetic state $\mathbf{s}^{'}_t$. After executing $a_t$, the next state $\mathbf{s}^{'}_{t+1}$ and reward $r_t$ are observed. The goal is for the agent to maximize cumulative future rewards, discounted by a factor $\gamma$ at each time step. The future discounted return at time $t$ is $ R_t = \sum_{t^{'}=t}^T\gamma^{t^{'}-t}r_{t^{'}}$, where $T$ is the termination time. Using the Bellman equation, the action-value function $Q(s^{'}_t,a_t)$ is
\begin{equation}
    Q(s^{'}_t,a_t) = \mathbb{E} \left[ r_t + \gamma \max_{a_{t+1}} Q(s^{'}_{t+1}, a_{t+1}) \mid s^{'}, a \right].
\end{equation}
In DQN, Q values for all actions are generated by a MLP where the input is the state $s^{'}_t$ and the output $\mathbf{Q}(s^{'}_t) \in \mathds{R}^{10}$ represents the possible actions for $a^q_t$ and $a_t^u$. Initially set as $Q(s^{'},a|\vartheta)$, the network is updated iteratively using a minibatch of experiences sampled from a fixed-capacity replay memory. For each experience in the minibatch, the target Q values $y_t$ are computed using the Bellman equation which is
\begin{equation}
    y_t=r_t+\gamma \max_{a_{t+1}} Q(s^{'}_{t+1},a_{t+1};\bar{\vartheta})
\end{equation}
where $\bar{\vartheta}$ is the target network parameters. The policy network update is performed by minimizing the loss function
\begin{equation}
    L(\vartheta)= \mathbb{E}_{(s^{'}_t,a_t,r_t,s^{'}_{t+1})\sim \text{replay buffer}}[(y_t-Q(s^{'}_t,a_t;\vartheta))^2].
\end{equation}
After each backpropagation of $Q(s^{'}_t,a_t;\vartheta)$, the target network $Q(s^{'}_t,a_t;\bar{\vartheta})$ is updated incrementally with a small fraction of the policy network. In specific,
\begin{equation}
    \bar{\vartheta}\gets \tau \vartheta +(1-\tau)\bar{\vartheta},
\end{equation}
where $\tau\ll 1$ \cite{sun2025out,vaghefi2024additive,wang2025machining,ogoke2021thermal,zhang2025employing,mnih2013playing,liu2018effects}. Once the policy network provides Q-values for all possible actions, the agent selects the actions using an $\epsilon$-greedy policy in an asynchronous manner. An example of synthetic actions in flow rate and temperature executed asynchronously with $\lambda=10$ is illustrated in Fig. \ref{Fig3.deepQ_method}(b). The corresponding synthetic vision distributions is shown in Fig. \ref{Fig3.deepQ_method}(c). The process begins with a flow rate of 191\% and a temperature of 213.3$^\circ$C, eventually converging to a flow rate of 101\% and a temperature of 210.0$^\circ$C.

\subsubsection{Four-phase training}\label{Reward shaping for improved convergence}

While synthetic transitions offer flexibility in adjusting DQN hyperparameters, designing an effective reward function and achieving stable convergence remain significant challenges—key bottlenecks in RL training. A major difficulty arises from the coupling effect between flow rate and temperature: different parameter combinations can yield similar extrusion volumes. For example, a higher temperature combined with a lower flow rate can produce results comparable to those from a lower temperature with a higher flow rate. To address this, the reward function is designed to decrease as the distance from the center of an ellipse increases, with the ellipse representing the optimal balance between flow rate and temperature. This elliptical reward structure reflects the parameters interdependence, encouraging the agent to favor combinations that optimize extrusion quality. The ellipse is oriented to prioritize favorable trade-offs, rewarding higher temperatures paired with lower flow rates and vice versa. Specifically, the reward function is
\begin{equation}
    r_t = 2/(1+\sqrt{(\bar{q}_t^2/a^2+\bar{u}_t^2/b^2)}) -1,\label{eq:reward}
\end{equation}
with
\begin{equation}
    \bar{q}_t = (q_t-q^*)\cos{\vartheta}-(u_t-u^*)\sin{\vartheta},
\end{equation}
\begin{equation}
    \bar{u}_t = (q_t-q^*)\sin{\vartheta}+(u_t-u^*)\cos{\vartheta},
\end{equation}
where $a$ and $b$ are adjustable major and minor axes of the ellipse. $q^*$ and $u^*$ are respectively the optimal flow rate and temperature. $\vartheta$ is the rotation angle.

In \eqref{eq:reward}, rewards are scaled within the range of -1 to 1 to stabilize the training process and ensure smooth transitions during reward shaping across the first three training phases \cite{schulman2015trust,mnih2013playing,mnih2016asynchronous}. This scaling prevents abrupt changes in learning dynamics, promoting steady convergence and reducing the risk of unstable updates. Throughout the initial three phases, the simulation environment assigns the class with the highest probability to correspond with the correct classification outcome. In specific, the extrusion class $x_0$ in the synthetic state (as defined in \eqref{eq:hat_pi}) is designated as the ground truth. This setup enables the controller to develop effective responses under ideal conditions while accounting for the inherent variability of the 3D printing process, with a primary focus on achieving stable convergence through reward shaping. In this context, reward shaping involves gradually tightening the reward function by sequentially halving the lengths of the ellipse’s minor and major axes. This progressive contraction of the reward space narrows the acceptable range for parameter combinations, encouraging the agent to make increasingly precise adjustments as training advances. The effectiveness of this three-stage reward shaping process is illustrated in Fig. \ref{Fig3.deepQ_method}(d). By progressively increasing the stringency of reward criteria, this structured method allows the policy network to adapt incrementally to stricter conditions, ultimately converging toward the optimal set of flow rate and temperature parameters.

In the fourth phase, the DQN is exposed to scenarios involving a controlled degree of misclassification, simulating the inevitable inaccuracies of the vision system. Specifically, $x_0$ is randomly selected from the two incorrect classes with probability $\varrho$, corresponding to the vision system’s precision level, defined by the top-1 accuracy of the vision classification task. This exposure trains the controller to manage real-world imperfections with resilience and adaptability. As a result, the agent learns to exercise appropriate hesitation in uncertain situations, improving its ability to make reliable, real-time adjustments during deployment.

\subsection{Zero-shot 3D printer deployment}\label{Zero-shot 3D printer deployment}

\begin{figure}[!t]
\centerline{\includegraphics[width=\columnwidth]{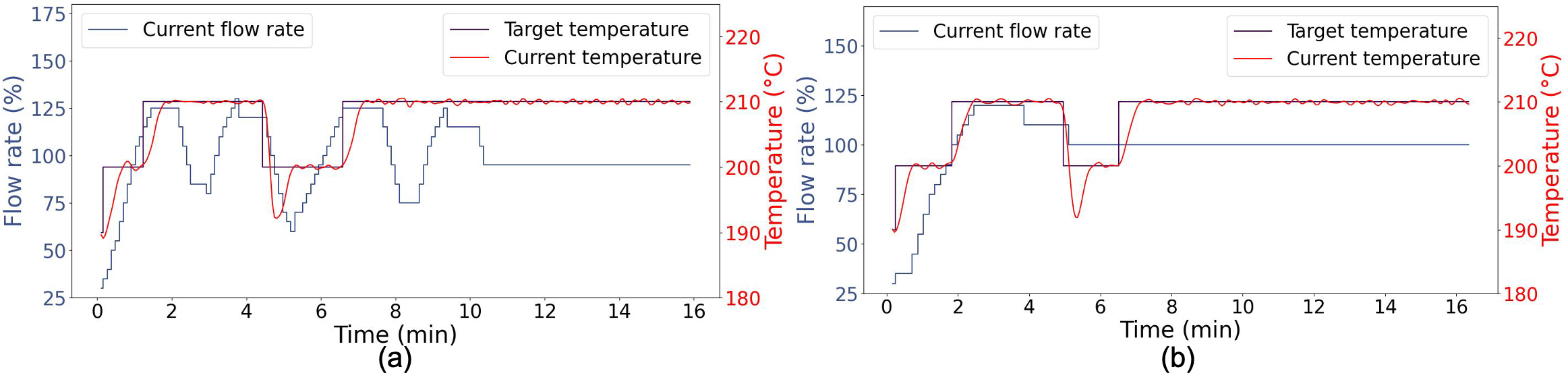}}
\caption{Showcase of under-extrusion error at 30\% flow rate and 190$^\circ$C nozzle temperature: (a) 3s waiting time; (b) 6s waiting time.}
\label{Fig3.zero_shot_deployment}
\end{figure}

The agent, which integrates uncertainty quantification with RL control, is directly deployed on a real 3D printer without requiring further online physical interaction. Effective control depends on capturing and analyzing an image sequence that accurately represents each printing segment. Two critical parameters govern this process: window size, which determines the number of images analyzed before making a control decision, and waiting time, which specifies the delay before capturing images following an action. After sufficient tests using a 20 Hz endoscope, it is revealed that a window size of 20 images strikes the best balance between vision precision and data processing efficiency. Meanwhile, the waiting time is optimized to 6 seconds to maintain a stable balance between control efficiency and the representativeness of the captured images. This delay is critical, as the extrusion state immediately after an action often undergoes transient fluctuations. Capturing images during this unstable period can introduce variability, potentially confusing the controller and degrading overall performance. A 6-second waiting period allows the extrusion process to stabilize, providing consistent and reliable data for decision-making. As shown in Fig. \ref{Fig3.zero_shot_deployment}, a comparison between 3-second and 6-second waiting times highlights the importance of this parameter. With a 3-second wait, the agent requires 95 steps to reach the optimal state, with an average duration of 6.44 seconds per action, totaling 10.21 minutes. In contrast, a 6-second wait reduces the number of steps to just 40, with an average duration of 9.43 seconds per action, leading to a faster total convergence time of 6.29 minutes. These results demonstrate that insufficient waiting time increases fluctuations and instability before convergence, ultimately prolonging the time needed to reach optimal control. This finding underscores the importance of incorporating both an adequately sized image window and a sufficient waiting period. Together, these parameters enhance control reliability and ensure the agent effectively manages dynamic transitions in real-time 3D printing operations.

\section{Results}

\subsection{Data generation and preprocessing}
Nozzle tip images are collected and labeled with corresponding flow rate and temperature values using the CAXTON system \cite{brion2022generalisable}. Flow rates are obtained directly from firmware settings, while temperature data include both real-time measurements and target values. The dataset is generated by printing multilayer cuboids through a bidirectional process. Since previous layers influence the visual quality of new layers, data collection occurs on every other layer. For non-collected layers, optimal parameters (100\% flow rate and 210$^\circ$C nozzle temperature for PLA) are applied. An endoscope positioned near the nozzle captures images at 20 Hz, ensuring consistent stability. Preprocessing isolates key areas corresponding to newly extruded lines, typically 87 pixels long (one-fourth of the image size) and 20 pixels wide, which are resized to 48$\times$16 pixels. Images with clearly visible line thickness are marked as valid detections, while those with unclear lines or deviations are labeled as incorrect. Expert evaluation of 69,668 images results in 56,225 valid detections, yielding an accuracy of 80.70\%. For training, 300 images are randomly selected for each of the 15 flow rate and temperature combinations (three flow rate classes: $<$90\%, 90\%-110\%, and $>$110\%; and five temperatures: 190$^\circ$C, 200$^\circ$C, 210$^\circ$C, 220$^\circ$C, and 230$^\circ$C). This produces 4,500 training images, with the remaining 51,725 reserved for offline testing to evaluate the vision system’s performance.

\subsection{Vision perception}
\begin{figure*}[!t]
\centerline{\includegraphics[width=\textwidth]{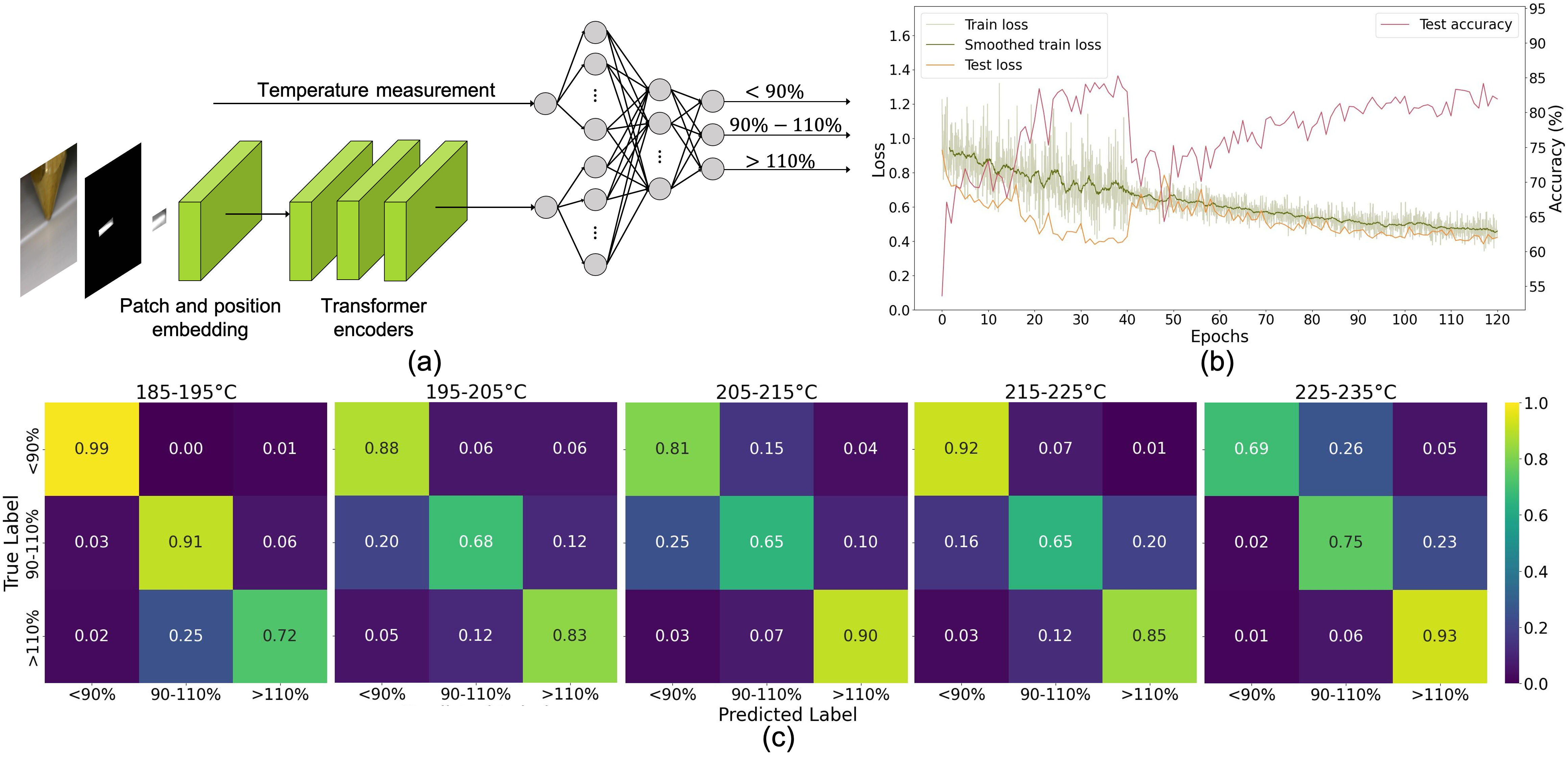}}
\caption{ViT training and testing: (a) ViT classification; (b) offline loss and accuracy trends; (c) confusion matrices at epoch 73.}
\label{Fig.CV_accuracy}
\end{figure*}
Using images categorized by flow rate class, the ViT-based model extracts features that are combined with real-time temperature measurements. These features are processed through classification layers, generating probability outputs for each flow rate class to assess the current extrusion condition. A cosine annealing learning rate schedule, starting at $10^{-4}$ with a full cycle spanning 100 epochs, optimizes training. The ViT’s architecture (Table \ref{Table:ViT_hyperparameters}) is fine-tuned to maximize accuracy on the offline test set. The training loss, test loss, and test accuracy progression with a batch size of 256 are shown in Fig. \ref{Fig.CV_accuracy}(b). 
\begin{table}[h!]
\centering
\caption{ViT configuration and parameter details}

\setlength{\tabcolsep}{6pt} 

\begin{tabular}{|c|c|c|c|c|c|}
\hline
Layers & Hidden dim & Patch size & Heads & MLP params & ViT params \\
\hline
3 & 256 & 4 & 8 & 33,363 & 4.0M \\
\hline
\end{tabular}

\label{Table:ViT_hyperparameters}
\end{table}
The smoothed training loss (window size of 30) decreases from 0.95 to 0.46. Despite fluctuations during the first 40 epochs, the model stabilizes and achieves robust generalization. Although the offline test accuracy reaches 82.37\% after 120 epochs, the model at 73 epochs achieving 78.58\% accuracy is selected for deployment. At 73 epochs, the model demonstrates a well-balanced trade-off between learning depth and generalization. The confusion matrices in Fig. \ref{Fig.CV_accuracy}(c) illustrate the model’s classification accuracy across the three flow rate classes and five temperature ranges. While the model demonstrates high offline accuracy, real-world deployment presents challenges due to dynamic process conditions and out-of-distribution data. Through experimentation, an image window size of 20 and a 6-second waiting period are determined as optimal to minimize classification errors and ensure reliable decision-making. The model exhibits strong extrapolation capabilities by effectively linking extrusion conditions to line thickness. It learns to associate thin lines with insufficient extrusion and thick lines with excessive material deposition. This understanding enables the model to generalize beyond the training data, maintaining high classification accuracy, often exceeding 90\%, even for flow rates outside the trained range.

\subsection{Deep Q-learning controller}

\begin{figure}[!t]
\centerline{\includegraphics[width=1\textwidth]{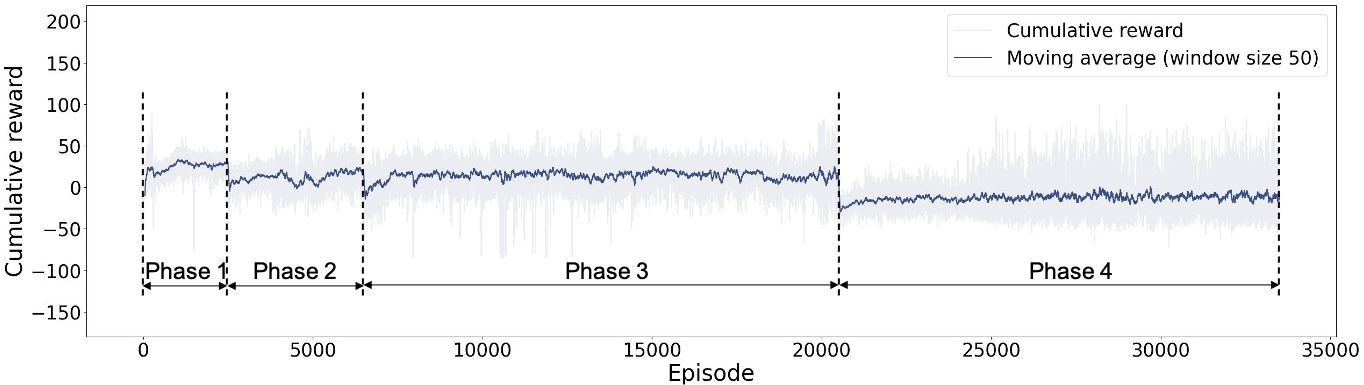}}
\caption{Cumulative reward per episode during four-phase training: reward shaping and 70\% simulated accuracy. }
\label{fig:Cumulative reward}
\end{figure}

\begin{figure}[!t]
\centerline{\includegraphics[width=0.9\textwidth]{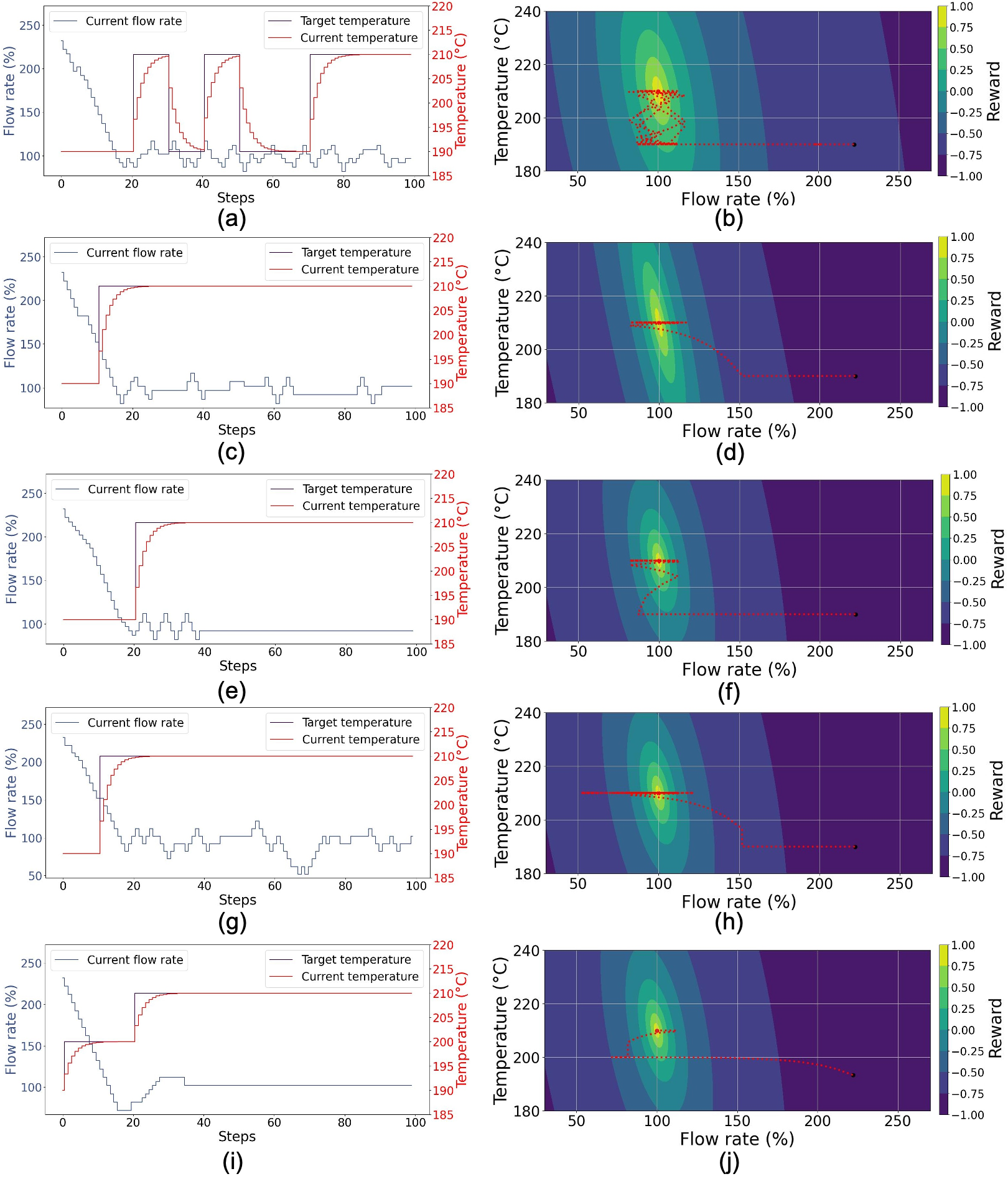}}
\caption{Agent performance in reward shaping: (a) actions in phase 1; (b) convergence plot of (a); (c) actions in phase 2; (d) convergence plot of (c); (e) actions in phase 3; (f) convergence plot of (e). Agent performance under 70\% classification accuracy: (g) actions of agent trained in phase 3; (h) convergence plot of (g); (i) actions of agent trained in phase 4; (j) convergence plot of (i). }
\label{fig:Agent performance in three phases of reward shaping}
\end{figure}

The selection of network architecture and the number of previous classes $\eta$ in the state representation are crucial for mitigating the effects of randomness and inaccuracies in vision perception and temperatures. The initial lenient reward function and synthetic transitions identify the most efficient network architecture for deep Q-learning. After numerous testing, $\eta$ was optimized to 30. The DQN architecture, implemented as a three-layer MLP with 47,082 parameters, processes 33 vision perception results and 2 temperature inputs to generate Q-values for action selection. The first hidden layer maps vision inputs to 256 features and temperature inputs to 32 features, which are concatenated. The second hidden layer reduces these 288 features to 128, and the output layer maps them to 5 Q-values each for the flow rate and temperature actions. Other hyperparameters are shown in table \ref{Table:deep_q_learning_hyperparameters}. Fig. \ref{fig:Cumulative reward} illustrates the cumulative reward per episode across the four training phases, with the first three involving gradually stricter reward shaping. During this process, cumulative rewards initially decrease as the reward function tightens.
\begin{table}[h!]
\centering
\caption{Deep Q-learning hyperparameters}

\setlength{\tabcolsep}{6pt} 

\begin{tabular}{|c|c|c|}
\hline
Description & Symbol & Value \\
\hline
The discount factor of future reward & $\gamma$ & 0.99  \\
Soft update factor & $\tau$ & 0.005  \\
Action frequency ratio & $\lambda$ & 10 \\
Batch size & - & 512  \\
Learning rate & - & $10^{-4}$ \\
Episode length & - & 100 steps \\
Ellipse semi-major axis & $a$ & 40 \\
Ellipse semi-minor axis & $b$ & 20 \\
Ellipse rotation angle & $\theta$ & 70$^\circ$ \\
\hline
\end{tabular}

\label{Table:deep_q_learning_hyperparameters}
\end{table}
The reduction occurs because tighter ellipses have a smaller high-reward area, making it harder for the agent, trained in previous phases, to achieve high rewards initially. Agent convergence is significantly enhanced due to the smaller reward area. For example, when an extrusion error occurs with a 232\% flow rate at 190$^\circ$C, the agent performance across the three phases of reward shaping is shown in Fig. \ref{fig:Agent performance in three phases of reward shaping}, subfigures (a, b), (c, d), and (e, f). These figures illustrate the progressive improvement as the agent adapts to error conditions. In the first phase, the agent converges but experiences fluctuations in both parameters. In the second phase, with the minor axis halved, temperature fluctuations are eliminated, though flow rate fluctuations persist. In the third phase, with the major axis halved, the agent converges without fluctuations in either parameter. This progression demonstrates the effectiveness of reward shaping in enhancing stability and precision in managing flow rate and temperature adjustments, even with significant parameter errors.

Up to this point, image classification results are based on the ground truth flow rate, assuming 100\% accuracy. However, this creates a gap between the simulation and real 3D printing. To reduce this discrepancy, the ground truth is selected with probability $\varrho$; otherwise, the class is randomly chosen from the other two options. The probability $\varrho$ can be adjusted to match the vision perception accuracy. As shown in Fig. \ref{fig:Agent performance in three phases of reward shaping}(g, h), when trained with reward shaping, the agent exposed to a stochastic environment with 70\% classification accuracy shows disturbances in flow rate actions. To address this, a fourth phase of offline training is introduced, where the agent interacts with an environment simulating inaccuracy with $\varrho=70\%$. As shown in Fig. \ref{fig:Agent performance in three phases of reward shaping}(i, j), this phase allows the agent to learn to handle classification inaccuracies and maintain good convergence. By the end, the agent is robust enough for real-world deployment, effectively managing flow rate and temperature adjustments despite vision perception inaccuracies.

\subsection{Real-world validation}

To validate the performance of the setpoint automation in addressing various extrusion errors, experiments were conducted with seven different flow rate values (30\%, 60\%, 80\%, 120\%, 150\%, 200\%, and 300\%) and three different temperatures (190$^\circ$C, 210$^\circ$C, and 230$^\circ$C). This section presents two specific cases of severe extrusion errors: case 1 involves over-extrusion at a 300\% flow rate and 230$^\circ$C temperature, while case 2 addresses under-extrusion at a 30\% flow rate and 190$^\circ$C temperature, as illustrated in Fig. \ref{Fig.real_experiment_plots}. The remaining tests are illustrated in appendix \ref{Setpoint adjustments from diverse extrusion error start points}. 
\begin{figure*}[!t]
\centerline{\includegraphics[width=0.9\textwidth]{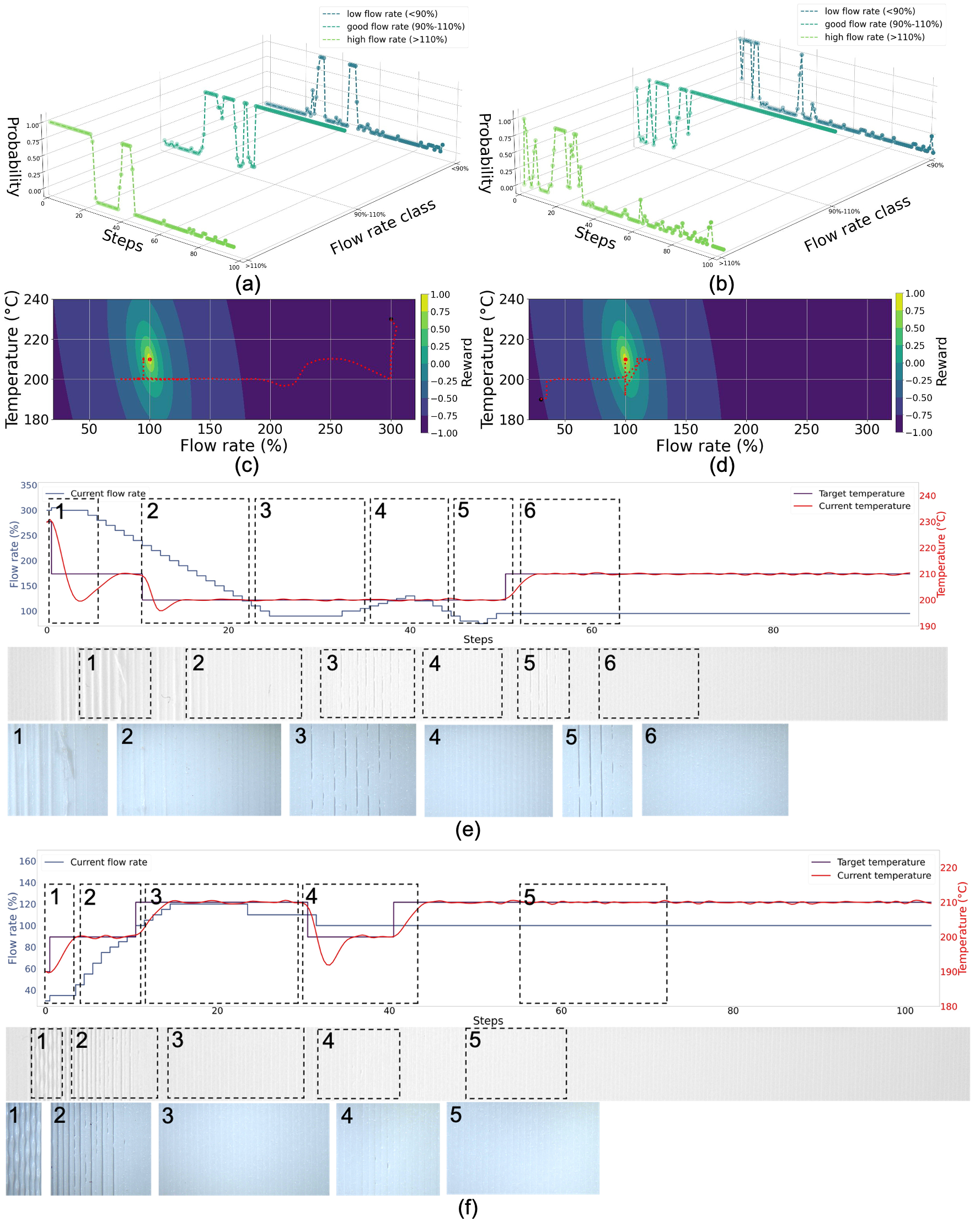}}
\caption{Error detection and correction in 3D printing: case 1 (over-extrusion: 300\%, 230$^\circ$C) (a, c, e) and case 2 (under-extrusion: 30\%, 190$^\circ$C) (b, d, f). Flow rate classification probabilities  (a, b); convergence plots (c, d); setpoint adjustments: parameter curves, printed samples, and microscope images (e, f).}
\label{Fig.real_experiment_plots}
\end{figure*}
The vision perception distributions shown in Fig. \ref{Fig.real_experiment_plots}(a) and \ref{Fig.real_experiment_plots}(b) indicate that when the printing is stable, the probability of misclassifying extrusion condition remains low, and transitions between different flow rate classes are smoother compared to the synthetic probabilities observed in simulation (Fig. \ref{Fig3.deepQ_method}(c)). This suggests that the DQN, trained in a simulated environment, operates under more challenging conditions compared to a real 3D printing process in terms of making accurate decisions based on flow rate classification probabilities. The parameter adjustments are plotted on the reward surface to depict the convergence behavior, as shown in Fig. \ref{Fig.real_experiment_plots}(c) and \ref{Fig.real_experiment_plots}(d). The printing parameters initially begin from the emulated over-extrusion or under-extrusion errors and subsequently converge, stabilizing near the optimal point. These plots provide a visual representation of how the controller navigates the reward landscape to achieve optimal process parameters. Fig. \ref{Fig.real_experiment_plots}(e) and \ref{Fig.real_experiment_plots}(f) illustrate the fluctuations in flow rate and temperature throughout the control process, along with the corresponding printed samples that reflect these adjustments. These figures provide a comprehensive view of how the dynamic parameter changes impact the quality of the printed outcomes. In case 1, severe over-extrusion, as depicted in the images within block 1 of Fig. \ref{Fig.real_experiment_plots}(e), is emulated by setting the flow rate to 300\% and the nozzle temperature to 230$^\circ$C. Initially, the flow rate shows a continuous decrease, as observed in block 2. The flow rate then undergoes two undershoots and one overshoot before stabilizing at the setpoint of 95\% shown in block 6. Specifically, the flow rate falls below the target by 5.26\% (reaching 90\%) in block 3 and by 21.05\% (reaching 75\%) in block 5 during the undershoots, while the overshoot in block 4 causes the flow rate to rise 36.84\% above the setpoint, reaching 130\%. Additionally, the temperature control experiences an undershoot between blocks 2 and 5, where the temperature briefly drops to 200$^\circ$C before stabilizing at the desired setpoint of 210$^\circ$C. In Fig. \ref{Fig.real_experiment_plots}(f), severe under-extrusion is emulated by setting the flow rate to 30\% and the nozzle temperature to 190$^\circ$C, as depicted in block 1. Following an initial phase of continuous flow rate increase, shown in block 2, the system experiences a single overshoot, with the flow rate briefly reaching 130\%, as observed in block 3. This value exceeds the final setpoint of 100\% (as shown in block 5) by 30\%. Similarly, the temperature setpoint undergoes an undershoot, as indicated in block 4, where the temperature temporarily drops to 200$^\circ$C before eventually stabilizing at the desired setpoint. The results confirm the agent’s ability to effectively correct extrusion errors and steer the printing process toward optimal parameters. The consistent convergence of printing parameters underscores the system’s robustness and reliability in real-world applications. More importantly, these findings validate the integration of vision-based uncertainty quantification with RL control, demonstrating the agent’s capacity to adapt dynamically, manage process variability, and maintain high print quality even under real-time operational challenges

\section{Conclusion}
We developed an advanced closed-loop setpoint adjustment framework for 3D printing that incorporates uncertainty quantification of printing segments directly into the decision-making process. The system begins with a vision module that extracts key extrusion areas, trained with sufficient ADR to enhance robustness. These areas are then transformed into scaled probability distributions representing classification results, providing a quantified measure of uncertainty for each printing segment. The asynchronous RL controller is trained through a structured four-phase process, guided by a tilted elliptical reward function that captures the coupling effects between flow rate and temperature. In the first three phases, the controller converges in an ideal environment where all classifications are assumed correct, allowing it to learn optimal decision-making under ideal conditions. The fourth phase exposes the controller to classification inaccuracies aligned with the vision system’s precision level, equipping the agent with appropriate hesitation. This combined approach allows for zero-shot deployment on a 3D printer, minimizing the sim-to-real gap and enabling reliable, real-time management of uncertainties during the printing process. Furthermore, the vision system is adaptable to various image analysis tasks, while the two-dimensional elliptical reward function can be extended to higher-dimensional ellipsoids to incorporate additional process parameters without losing the interdependencies between them. This scalability ensures that the proposed framework remains efficient and uncertainty-aware across a broad range of AM processes.

While our control scheme offers significant advancements, some limitations remain. The camera must be positioned to directly face the extrusion thickness, limiting flexibility. An omnidirectional detection system could address this by using multiple cameras, switching views based on nozzle movements derived from G-code instructions. Direct interaction with 3D printers could theoretically further reduce the sim-to-real gap, but our findings suggest that the performance gains do not justify the substantial costs in labor, time, and materials. Moreover, physical interactions could introduce additional variability, potentially confusing the policy network and diminishing overall robustness. In contrast, zero-shot learning offers a more practical and efficient solution. It simplifies deployment, avoids the complexities and costs of real-world training, and still achieves strong performance without compromising reliability or adaptability.

\appendices

\section*{Acknowledgment}

This work was funded by the Engineering and Physical Sciences Research Council (EPSRC) under award EP/V062123/1. X.L. and S.W.P. gratefully acknowledge this support. For the purpose of open access, the authors have applied a Creative Commons Attribution (CC BY) license to any Author Accepted Manuscript version arising from this submission. S.W.P. is co-founder of Matta Labs, a spin-out developing AI-based software for the manufacturing industry.

\bibliographystyle{elsarticle-num}  

\appendices
\renewcommand{\thefigure}{A.\arabic{figure}} 
\setcounter{figure}{0} 

\section{Setpoint adjustments from diverse extrusion error start points}\label{Setpoint adjustments from diverse extrusion error start points}


\begin{figure}[htbp]
\centerline{\includegraphics[width=0.8\columnwidth]{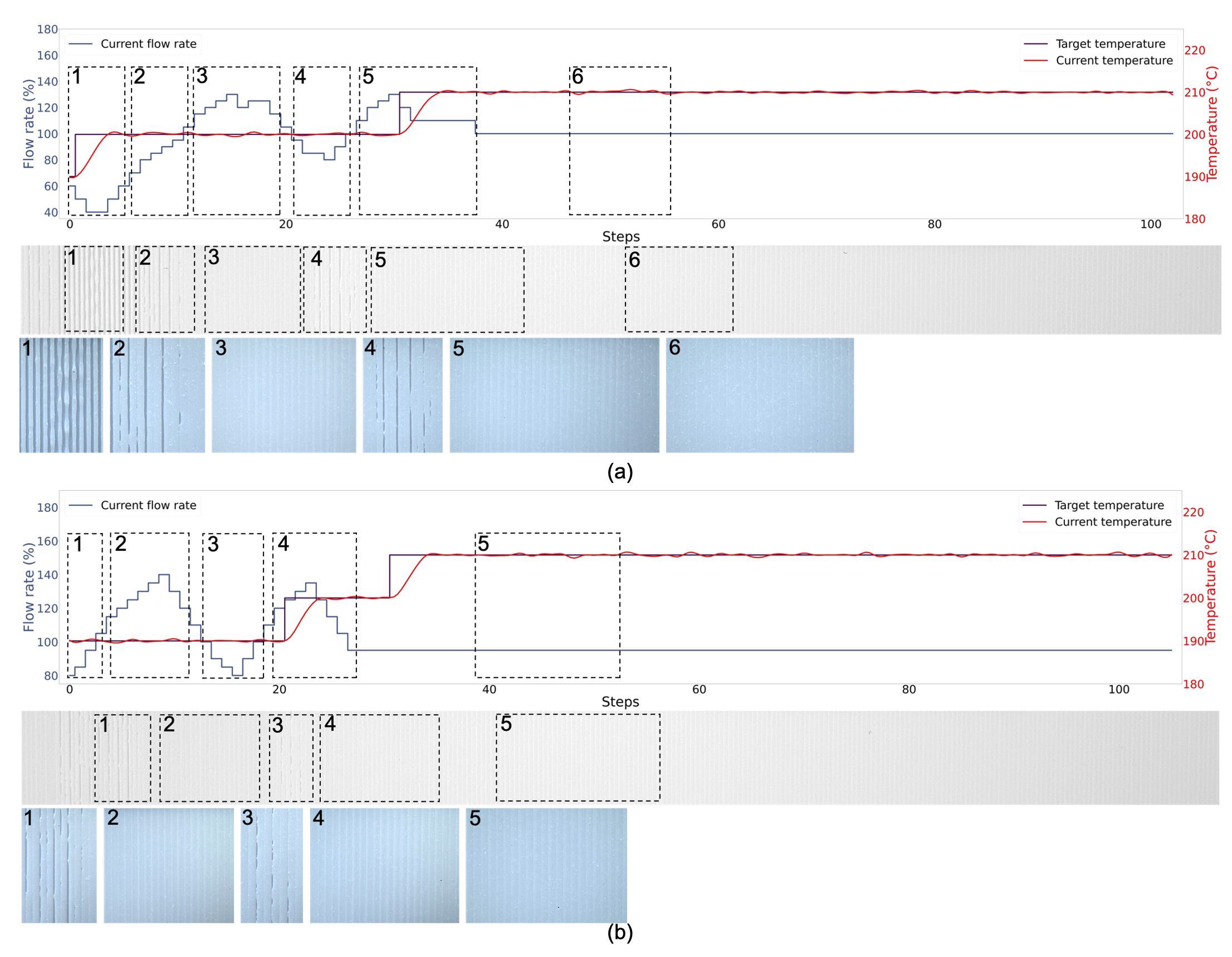}}
\caption{(a) 60\% flow rate and 190$^\circ$C nozzle temperature; (b) 80\% flow rate and 190$^\circ$C nozzle temperature.}
\label{Fig1.1_appendix}
\end{figure}

\begin{figure}[htbp]
\centerline{\includegraphics[width=0.8\columnwidth]{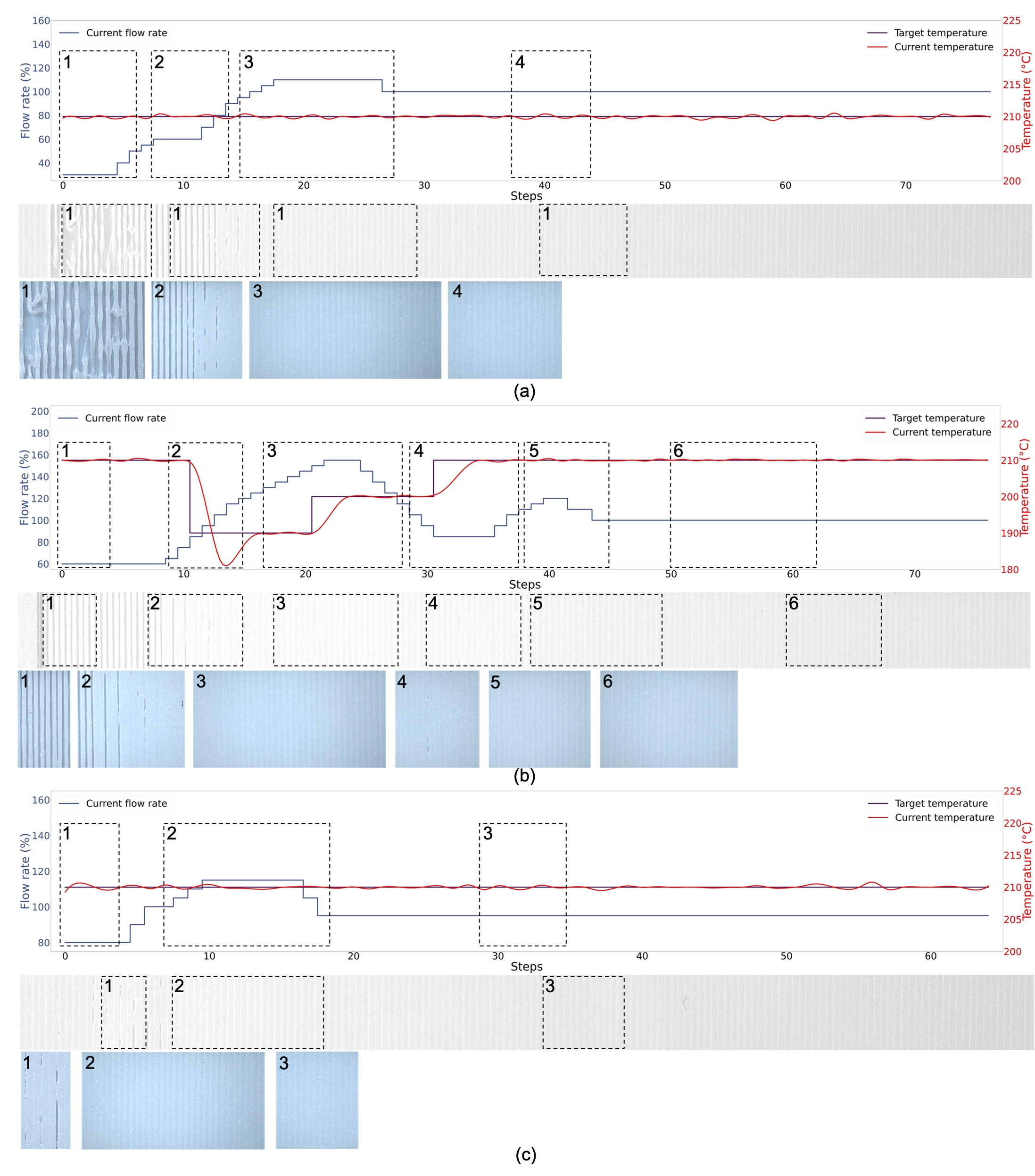}}
\caption{(a) 30\% flow rate and 210$^\circ$C nozzle temperature; (b) 60\% flow rate and 210$^\circ$C nozzle temperature; (c) 80\% flow rate and 210$^\circ$C nozzle temperature.}
\label{Fig1.2_appendix}
\end{figure}

\begin{figure}[htbp]
\centerline{\includegraphics[width=0.8\columnwidth]{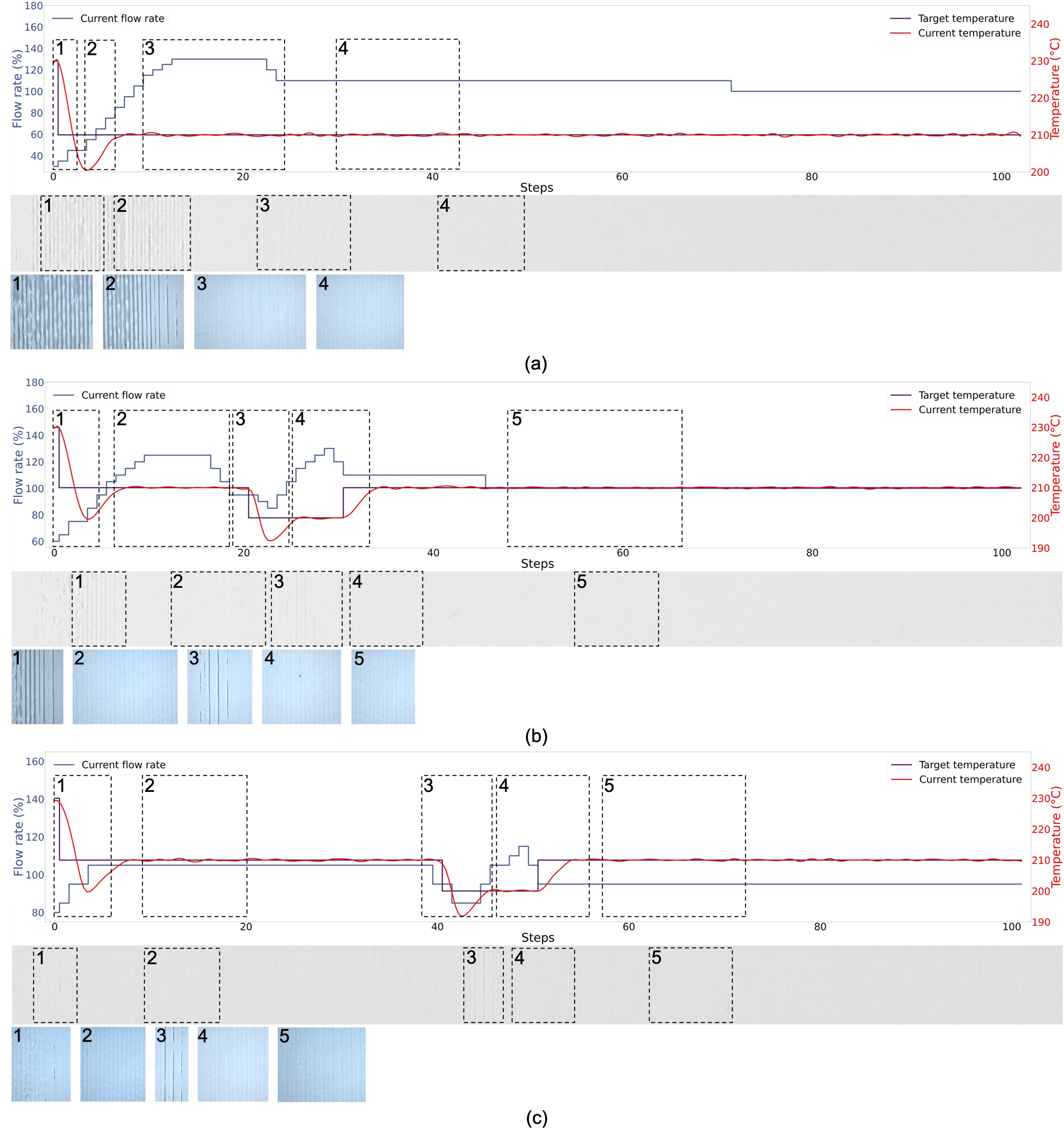}}
\caption{(a) 30\% flow rate and 230$^\circ$C nozzle temperature; (b) 60\% flow rate and 230$^\circ$C nozzle temperature; (c) 80\% flow rate and 230$^\circ$C nozzle temperature.}
\label{Fig1.3_appendix}
\end{figure}


\begin{figure}[htbp]
\centerline{\includegraphics[width=0.8\columnwidth]{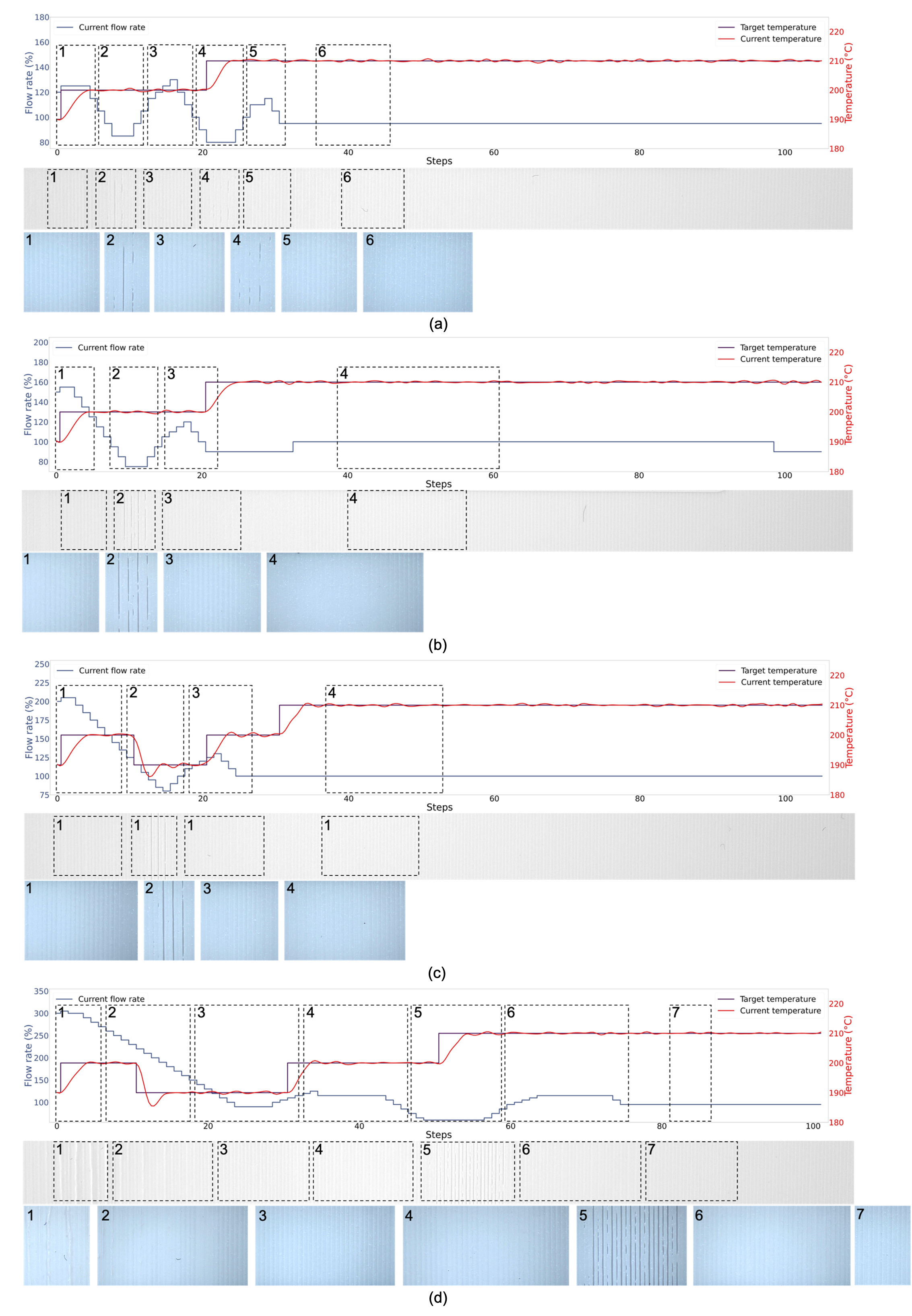}}
\caption{(a) 120\% flow rate and 190$^\circ$C nozzle temperature; (b) 150\% flow rate and 190$^\circ$C nozzle temperature; (c) 200\% flow rate and 190$^\circ$C nozzle temperature; (d) 300\% flow rate and 190$^\circ$C nozzle temperature.}
\label{Fig1.4_appendix}
\end{figure}

\begin{figure}[htbp]
\centerline{\includegraphics[width=0.8\columnwidth]{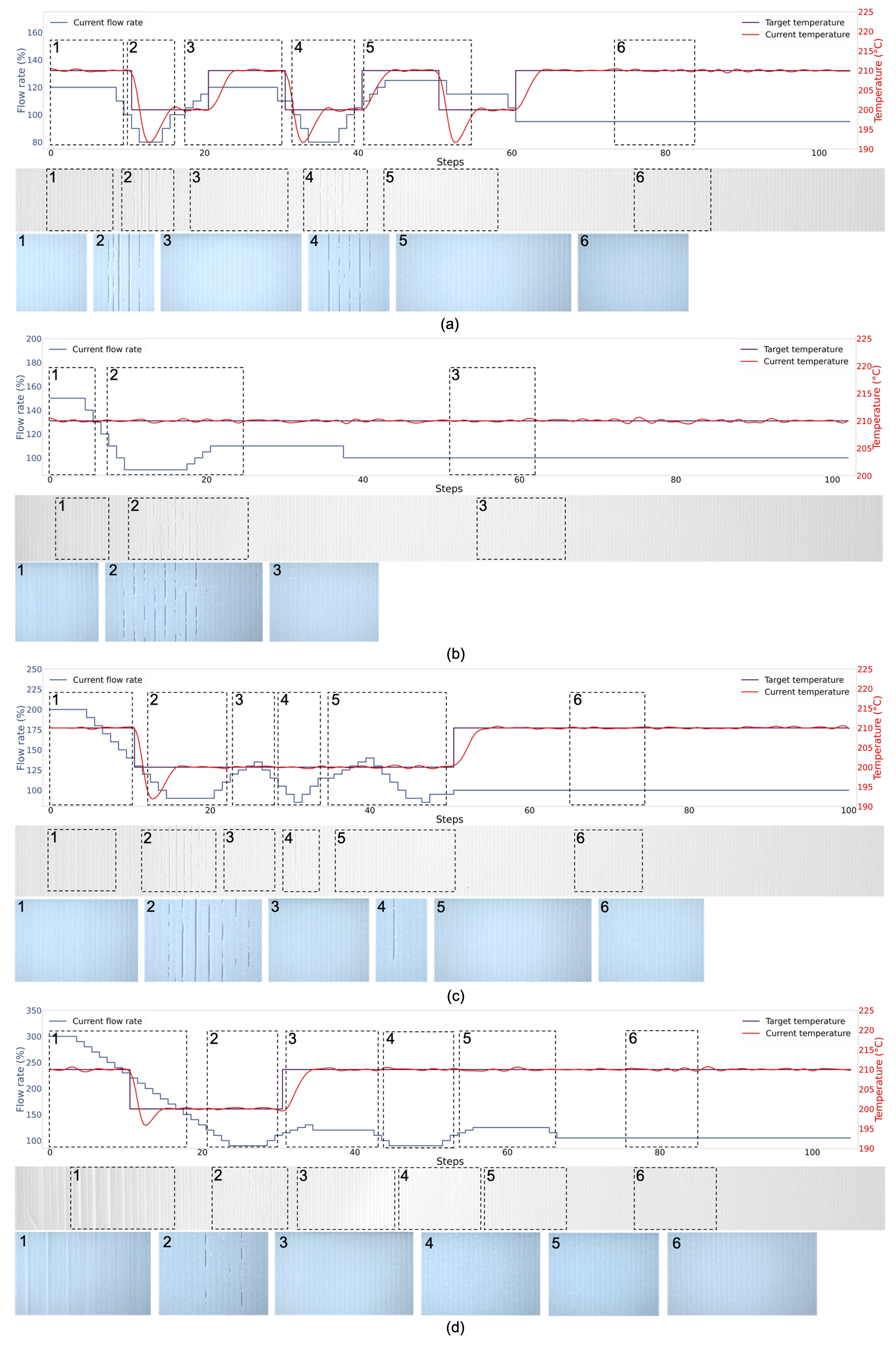}}
\caption{(a) 120\% flow rate and 210$^\circ$C nozzle temperature; (b) 150\% flow rate and 210$^\circ$C nozzle temperature; (c) 200\% flow rate and 210$^\circ$C nozzle temperature; (d) 300\% flow rate and 210$^\circ$C nozzle temperature. }
\label{Fig1.5_appendix}
\end{figure}

\begin{figure}[htbp]
\centerline{\includegraphics[width=0.8\columnwidth]{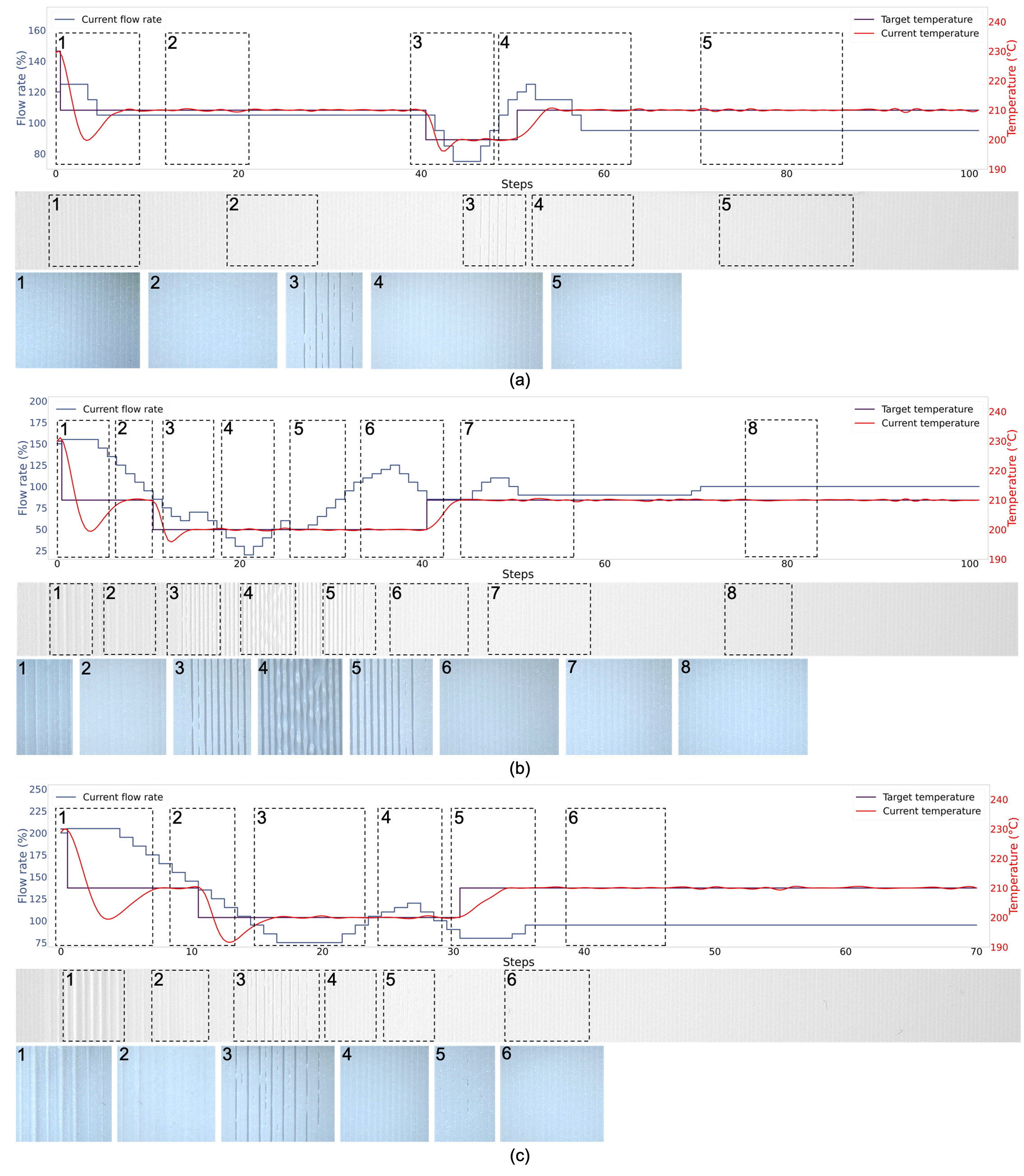}}
\caption{(a) 120\% flow rate and 230$^\circ$C nozzle temperature; (b) 150\% flow rate and 230$^\circ$C nozzle temperature; (c) 200\% flow rate and 230$^\circ$C nozzle temperature. }
\label{Fig1.6_appendix}
\end{figure}

\end{document}